\documentclass[a4paper,11pt]{article}
\usepackage{pos}
\usepackage{cleveref}

\newcommand{\chpt}{{$\chi$PT}}
\newcommand{\rcite}[1]{Ref.~\cite{#1}}
\newcommand{\rrcite}[1]{Refs.~\cite{#1}}

\renewcommand{\d}{\mathrm d}
\renewcommand{\Im}{\mathrm{Im}\,}
\renewcommand{\v}[1]{\boldsymbol{#1}}

\newcommand{\constr}[2]{\langle#1,#2\rangle}
\newcommand{\sat}{\mathcal B}
\newcommand{\rel}{\mathcal R}
\renewcommand{\mho}{\text{\raisebox{-.5pt}{\rotatebox[origin=c]{180}{$\Omega$}}}}

\usepackage{amsthm}
\newtheorem{proposition}{Proposition}

\usepackage{pgf,tikz}
\usepackage{pgfplots}
\pgfplotsset{compat=1.15}
\usepgfplotslibrary{fillbetween}
\usepgfplotslibrary{colorbrewer}
\usepgfplotslibrary{colormaps}
\usepgfplotslibrary{groupplots}
\usepgfplotslibrary{patchplots}
\newcommand{\hatchborderampl}{4pt}
\newcommand{\weakhatchborderampl}{3pt}
\definecolor{plotI}{HTML}{0077BB}
\definecolor{plotII}{HTML}{33BBEE}
\definecolor{plotIII}{HTML}{009988}
\definecolor{plotIV}{HTML}{EE7733}
\definecolor{plotV}{HTML}{CC3311}
\definecolor{plotVI}{HTML}{EE3377}
\definecolor{plotgray}{HTML}{BBBBBB}
\tikzset{
    hatchborder/.style={
        postaction={decorate, draw, thin}, decoration={border, segment length=2pt, amplitude=\hatchborderampl, angle=225}
    },
    weakhatchborder/.style={
        postaction={decorate, draw, thin}, decoration={border, segment length=3pt, amplitude=\weakhatchborderampl, angle=225}
    },
    invhatchborder/.style={
        postaction={decorate, draw, thin}, decoration={border, segment length=2pt, amplitude=\hatchborderampl, angle=45}
    },
    weakinvhatchborder/.style={
        postaction={decorate, draw, thin}, decoration={border, segment length=3pt, amplitude=\weakhatchborderampl, angle=135}
    },
    relevant/.style={very thick, hatchborder},
    irrelevant/.style={thin, opacity=0.5, weakhatchborder},
    facet/.style={color=., fill=., fill opacity=#1, join=round}, facet/.default={.8},
    axes to fid/.style={thick, black, densely dotted},
    surface to fid/.style={thick, black, stealth-}
}

\usetikzlibrary{external}
\title{NNLO Positivity Bounds on \chpt\\for a General Number of Flavours}
\ShortTitle{NNLO Positivity Bounds on \chpt}

\author[a]{Benjamin Alvarez}
\author[b]{Johan Bijnens}
\author*[b]{Mattias Sjö}

\affiliation[a]{Aix Marseille Université, Université de Toulon,\\
  CNRS, CPT, Marseille, France}

\affiliation[b]{Department of Astronomy and Theoretical Physics, Lund University,\\
  Box 43, SE 22100 Lund, Sweden\\~}

\emailAdd{benjamin-alvarez@univ-tln.fr}
\emailAdd{bijnens@thep.lu.se}
\emailAdd{mattias.sjo@thep.lu.se}

\abstract{We present positivity bounds, derived from the principles of analyticity, unitarity and crossing symmetry, that constrain the low-energy constants of chiral perturbation theory.
Bounds are produced for 2, 3 or more flavours with equal meson masses, up to and including next-to-next-to-leading order (NNLO), using the second and higher derivatives of the amplitude. We enhance the bounds by using the most general isospin combinations posible (or higher-flavour counterparts thereof) and by analytically integrating the low-energy range of the amplitude. In addition, we present a powerful and general mathematical framework for efficiently managing large numbers of positivity bounds.}

\FullConference{%
  The 10th International Workshop on Chiral Dynamics - CD2021\\
  15-19 November 2021\\
  Online
  }

\renewcommand{\logo}{\relax}

\begin{document}
\maketitle

Chiral Perturbation Theory (\chpt) is the most widespread effective field theory (EFT) for low-energy QCD.
However, its predictive power is limited by the number of free parameters (low-energy constants, LECs) in its Lagrangian \cite{Gasser:1983yg,Gasser:1984gg,Bijnens:1999sh,Bijnens:2018lez}: Even neglecting non-strong interactions, 102 LECs appear up to NNLO in the low-energy expansion, corresponding to two-loop amplitudes, with another 1233 entering at the next order.
Not all of these feature in amplitudes of interest, but the fact remains that the number of LECs limits the usefulness of higher-order \chpt\ corrections to observables.
Only the LO LECs, corresponding to the pion mass and decay constant, are known to high precision; the NLO LECs are known at the percent level, only educated guesses are available at NNLO, and nothing at all at N$^3$LO \cite{Bijnens:2014lea}.

A possible mitigation of this issue comes from the fact that, besides measurements in experiments and on the lattice, it is possible to constrain the values of the LECs also from the purely theoretical side.
All quantum field theories must obey the principles of analyticity, unitarity and crossing symmetry, but it turns out that these are not necessarily compatible with the assumption of perturbativity for an EFT; thus, imposing all four principles can lead to non-trivial requirements on the Lagrangian.
This concept was pioneered by Martin \cite{Martin:1969ina} before the development of \chpt\ as such, with renewed interest in recent decades \cite{Pham:1985cr,Ananthanarayan:1994hf,Pennington:1994kc,Dita:1998mh}.
Our work, published in full as \rcite{Alvarez:2021kpq}, is based on the methods of Manohar \&\ Mateu \cite{Manohar:2008tc,Mateu:2008gv} with further inspiration from \rrcite{Wang:2020jxr,Tolley:2020gtv}; other recent work in similar directions includes \rrcite{Bellazzini:2020cot,Caron-Huot:2020cmc,Sinha:2020win}.

\section{Positivity Bounds}
Throughout, we shall work in the isospin limit (all mesons having the same mass, $M$) and use the normalized Mandelstam variables $s=(p_1+p_2)^2/M^2$, $t=(p_1+p_3)^2/M^2$ and $u=(p_1+p_4)^2/M^2$.

Following Manohar \&\ Mateu, the process of obtaining bounds starts with the isospin decomposition of the $2\to2$ pseudoscalar meson scattering amplitude,
\begin{equation}
    T(s,t) = a_J T^J(s,t)\,,
\end{equation}
implicitly summed over the label $J$, which for two-flavour \chpt\ runs over isospin channels $0,1$ and $2$; with three flavours, this generalizes to the five representation labels $I,A,S,AS$ and $SS$, with a sixth, $AA$, appearing in the unphysical case of four or more flavours.
With this decomposition, $s\to u$ crossing symmetry is implemented as
\begin{equation}
    T^I(u,t) = C^{IJ}_u T^J(s,t)\,,
\end{equation}
with the matrix $C^{IJ}_u$ determined entirely from the group structure.

Next, we invoke analyticity to write the $k$-times-subtracted fixed-$t$ dispersion relation,
\begin{equation}
    a_J\frac{\d^k}{\d s^k} T^J(s,t) = \frac{k!}{2\pi i}\oint \d z\:\frac{a_J T^J(z,t)}{(z-s)^{k+1}}\,,
\end{equation}
which through contour manipulation can be brought into the form
\begin{equation}\label{eq:integral}
    a_J\frac{\d^k}{\d s^k} T^J(s,t) = \frac{k!}{\pi}\int_4^\infty \d z\left[\frac{a_J}{(z-s)^{k+1}} + \frac{(-1)^k a_I C^{IJ}_u}{(z-u)^{k+1}}\right] \Im T^J(z+\varepsilon i,t)\,.
\end{equation}
The two terms in parentheses stem from routing the contour along the cuts corresponding to the $s$- and $u$-channel, respectively; with normalized Mandelstam variables, 4 corresponds to threshold.

Above threshold, and within a wide domain of convergence, we may partial-wave expand the amplitude as
\begin{equation}
    T^J(s,t) = \sum_{\ell=0}^\infty (2\ell+1)f_\ell^J(s) P_\ell\Big( 1 + \frac{2t}{s-4}\Big)\,,
\end{equation}
where $P_\ell$ are Legendre polynomials, and the optical theorem (invoking unitarity) imposes for the partial-wave amplitudes $f_\ell^J$ that
\begin{equation}
    \Im f_\ell^J(s) = s\sigma_\ell^J(s)\sqrt{1 - 4/s}\,,
\end{equation}
which is positive above threshold since the partial-wave cross-sections $\sigma_\ell^J(s)$ are.
Thus, in the range where $P_\ell\Big( 1 + \frac{2t}{s-4}\Big)$ is positive, $\Im T^J(s,t)$ must be positive as well.
Putting all of this together, we find
\begin{subequations}\label{eq:main}
    \begin{align}
        a_J\frac{\d^k}{\d s^k}T^J(s,t) &\geq 0 \label{eq:main:constr}\\ 
        \text{if}\quad a_I\left\{\delta^{IJ}\left[\frac{z-u}{z-s}\right]^{k+1} + (-1)^k C_u^{IJ}\right\} &\geq 0\quad\text{for all $z\geq 4$ and all $J$}\,,\label{eq:main:cond}
    \end{align}
\end{subequations}
valid in the range $t\in[0,4],s\in[-t,4]$, a below-threshold region free of singularities.
As follows from the Froissart bound \cite{Froissart:1961ux}, $k\geq 2$ is necessary and sufficient for convergence.
It furthermore turns out that odd $k$ are useless with three or more flavours, and forbidden with two.
Likewise, it can be shown that it suffices to satisfy \cref{eq:main:cond} at $z=4$ (threshold) and in the limit $z\to\infty$.

\Cref{eq:main} provides the means for producing positivity bounds by evaluating $a_J\d^kT^J(s,t)/\d s^k$ at fixed $s,t$ in the region of validity.
Conventionally, this is done with $a_J$ fixed to one of the mass eigenstates --- $\begin{pmatrix}0&0&1\end{pmatrix}$ for two-flavour $\pi^+\pi^+$ scattering, $\begin{pmatrix}   0&\tfrac15&0&\tfrac12&\tfrac3{10}\end{pmatrix}$ for three-flavour $\pi\eta$ scattering, etc.\ --- but in the isospin limit, this is not necessary.
The region in $a_J$-space that satisfies \cref{eq:main:cond} is rather broad and depends on $s,u$ --- there is no need to require validity for \emph{all} $s,u$, just at a fixed point --- and thus gives a wide range of bounds.

Inspired by the approach taken in \cite{Wang:2020jxr}, one may explicitly evaluate the lowest portion of the integral in \cref{eq:integral}, from $4$ up to some $\lambda$, and subtract it from the equation before applying the positivity arguments to the right-hand side.
This subtracts a known, positive quantity from the left-hand side of \cref{eq:main:constr}, thus strengthening the bounds.
Alternatively, a broader choice of $s,u$ and $a_J$ becomes available, which produces new, possibly stronger bounds despite the subtracted quantity now being possibly negative.
We have analytically performed the pertinent integral applied to the NNLO $2\to2$ $n$-flavour scattering amplitude \cite{Bijnens:2011fm} by integrating a wider class of functions that includes those appearing in the 1- and 2-loop equal-mass integrals, thus making this subtraction easy to perform.
Care has to be taken with the choice of $\lambda$: larger values strengthen the bounds, but since it is done at fixed order in the low-energy expansion, the validity decreases as $\lambda$ approaches the Chivukula--Dugan--Golden bound \cite{Chivukula:1992gi}, $\lambda \sim 70/n$, at which perturbative breakdown is expected.

\section{Linear Constraints}
Our generalization of $a_J$ beyond the mass eigenstates allows for the production of a practically unlimited number of independent bounds on the LECs, and our use of NNLO $n$-flavour \chpt\ greatly increases the dimension of the parameter space: The LECs appear in the bounds as up to 20 independent linear combinations, although this is reduced by using two- or three-flavour \chpt, by using higher derivative counts $k$, or by fixing $t=4$, which is the value at which most strong bounds are obtained.
Nevertheless, our bound-producing methods necessitate improved bound-managing methods; dissatisfied with those available in the literature, we have derived a new mathematical framework for this purpose.

Up to NNLO, $\d^k T(s,t)/\d s^k$ is an inhomogeneous linear function of the LECs; thus, the general expression of interest is of the form
\begin{equation}
    \alpha_1b_1 + \alpha_2 b_2 + \ldots + \alpha_N b_N = \v\alpha\cdot\v b \geq c\,,
\end{equation}
where $b_i$ are the values being constrained (here, the LECs) and $\alpha_i$ and $c$ are known [here, from  \cref{eq:main:constr}].
Thus, we introduce \emph{(linear) constraints}, denoted $\constr{\v\alpha}{c}$, and express the relation $\v\alpha\cdot\v b\geq c$ as `$\constr{\v\alpha}{c}$ is satisfied by $\v b$'.
We express the simultaneous application of multiple constraints as summation; thus, with
\begin{equation}
    \Omega = \sum_i \constr{\v\alpha_i}{c_i}\,,
\end{equation}
$\Omega$ is satisfied by $\v b$ if and only if $\v b$ satisfies all of $\constr{\v\alpha_i}{c_i}$.
We denote by $\sat(\Omega)$ the set of all points that satisfy $\Omega$.
Even if two constraints $\Omega$ and $\Omega'$ may be written as different sums, we consider them equal when $\sat(\Omega)=\sat(\Omega')$.

There is a natural notion of one constraint ($\Omega$) being stronger than another $(\Omega')$, which we write as $\Omega\geq\Omega'$ and define by $\sat(\Omega)\subseteq\sat(\Omega')$: Every point that satisfies $\Omega$ also satisfies $\Omega'$, but $\Omega$ may bring additional restrictions.
When generating many constraints, we wish to only retain the strongest among them and discard the rest.
Apart from some obvious identities such as 
\begin{equation}
    \begin{gathered}
        \Omega\leq(\Omega+\Omega')\geq\Omega',\qquad \constr{\v\alpha}{c} = \constr{\v\alpha/|c|}{\mathrm{sgn}(c)}\quad\text{if $c\neq 0$}\,,\\
        \constr{\v\alpha}{-1}\leq\constr{(1+\lambda)\v\alpha}{-1}\leq\constr{\v\alpha}{0}=\constr{\lambda\v\alpha}{0}\leq\constr{(1+\lambda)\v\alpha}{1}\leq\constr{\v\alpha}{1}\quad \text{if $\lambda\geq 0$}\,,
    \end{gathered}
\end{equation}
this is a highly nontrivial task, and is the focus of most of our effort.
The fundamental result, proven in \rcite{Alvarez:2021kpq}, is the following:
\begin{proposition}\label{prop:weaker}
    A linear constraint $\constr{\v\beta}{d}$ is weaker than the combined constraint $\Omega=\sum_i\constr{\v\alpha_i}{c_i}$, i.e.\ $\constr{\v\beta}{d}\leq\Omega$, if and only if there exist $\lambda_i\geq 0$ such that
    \begin{equation}
        \v\beta = \sum_i\lambda_i\v\alpha_i\,,\qquad \sum_i\lambda_i c_i\geq d\,.
    \end{equation}
\end{proposition}
\noindent This, however, is rather indirect, since no indication is given of how to find these $\lambda_i$.
A more direct result is the following:
\begin{proposition}\label{prop:construct}
    Given $\Omega$ as above, there exists $\mho_d=\sum_j\constr{\v n_j}{r_j}$ such that $\constr{\v\beta}{d}\leq\Omega$ if and only if $\v\beta\in\sat(\mho_d)$.
\end{proposition}
\noindent The constraint $\mho_d$ can be thought of as a dual of $\Omega$; in fact, applying \cref{prop:construct} to $\mho_d$ for $d\neq 0$ recovers $\Omega$ (Corollary~B.5 in \rcite{Alvarez:2021kpq}).
We provide a straightforward algorithm for finding $\constr{\v n_i}{r_i}$, which is stated in detail in appendix~B.4.1 of \rcite{Alvarez:2021kpq}.
In short, one forms a set consisting of certain linear combinations of the $\v\alpha_i$, and takes its \emph{convex hull} (the smallest convex set containing it), for which efficient algorithms exist \cite{qhull}.
As a side-effect, one obtains the normal vectors of the facets of the hull's surface, and after discarding certain facets based on some straightforward conditions, these normal vectors are essentially $\v n_j$.
Furthermore, there is a direct relation (Proposition~B.4 in \rcite{Alvarez:2021kpq}) between the $\v n_j$ and the location of the vertices and edges of $\sat(\Omega)$, which helps with visualization.

Lastly, we have the following:
\begin{proposition}\label{prop:relevant}
    Of all sets $\mathcal S$ such that $\Omega = \sum_{\constr{\v\alpha}{c}\in\mathcal S} \constr{\v\alpha}{c}$, there exists a smallest such set, denoted $\rel(\Omega)$.
    As long as $\Omega$ is non-degenerate, i.e.\ $\sat(\Omega)$ is not contained in any hyperplane, this smallest set is unique.    
\end{proposition}
\noindent Appendix~B.4.4 of \rcite{Alvarez:2021kpq} covers the algorithm for finding $\rel(\Omega)$, which essentially consists of retaining only those $\v\alpha_i$ that end up on the surface of the aforementioned convex hull.
Thus, we may generate as many constraints as we want, and \cref{prop:relevant} will pick out those that are actually relevant for placing bounds on the LECs.
One shortcoming is that some constraints only carve out a negligible corner of parameter space, while still being retained by the algorithms; we unfortunately do not have a systematic way of filtering out such `near-irrelevant' constraints.

\section{New Bounds on \chpt}
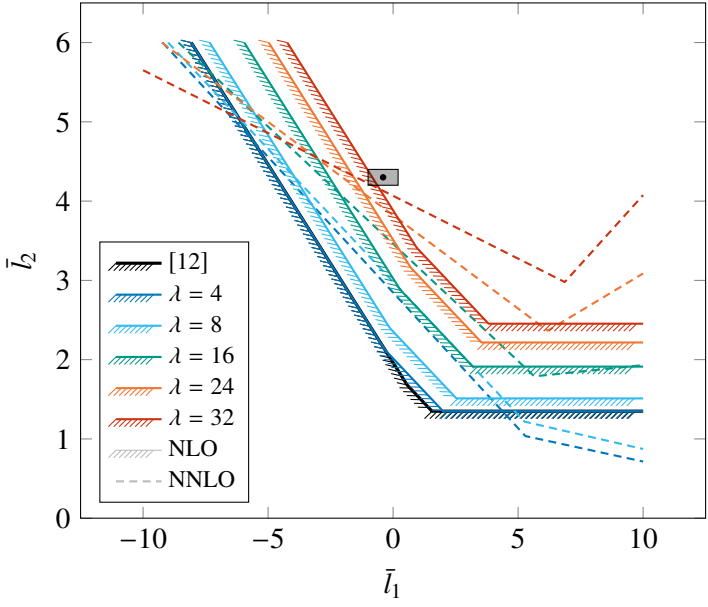
\begin{figure}[hbtp]
    \begin{minipage}[t]{0.65\textwidth}
        \raisebox{-\height}{
            \begin{tikzpicture}
                \begin{axis}[scale=1.2,
                    ylabel={$\bar l_2$}, xlabel={$\bar l_1$},
                    xmin=-12.5,xmax=12.5, ymin=0, ymax=6.5,
                    legend pos=south west, legend cell align=left,
                    legend image code/.code={
                        \draw (0cm,0cm) -- (0.6cm,0cm);
                    },
                    legend style={font=\footnotesize}]
                
                    \addlegendimage{black, very thick, hatchborder};
                    \addlegendimage{plotI,  thick, hatchborder};
                    \addlegendimage{plotII,  thick, hatchborder};
                    \addlegendimage{plotIII,  thick, hatchborder};
                    \addlegendimage{plotIV,  thick, hatchborder};
                    \addlegendimage{plotV,  thick, hatchborder};
                    \addlegendimage{plotgray,  hatchborder};
                    \addlegendimage{plotgray,  thick, densely dashed};

                    \draw[
                        black, join=round, very thick, hatchborder
                    ]  
                        (-8.075, 6)
                        -- (0.567,1.679)
                        -- (1.554,1.35)
                        -- (10,1.35);
                    
                    \draw[
                        plotI, join=round, thick, hatchborder
                    ]
                        (-8.07517, 5.99979)
                        -- (-0.324077, 2.12455)
                        -- (1.99952, 1.35006)
                        -- (9.9999, 1.35014);
                        \draw[
                        plotI, join=round, thick, densely dashed
                    ]
                        (-9.21773, 5.99961)
                        -- (5.2909, 1.03618)
                        -- (10.0005, 0.714619);
                        
                    \draw[
                        plotII, join=round, thick, hatchborder
                    ]
                        (-7.33653, 6.0002)
                        -- (-0.202206, 2.43291)
                        -- (-0.0944658, 2.38661)
                        -- (2.52571, 1.51313)
                        -- (2.54915, 1.51135)
                        -- (10.0003, 1.51097);
                              
                    \draw[
                        plotII, join=round, thick, densely dashed
                    ]  
                        (-9.00084, 6.0002)
                        -- (5.26049, 1.21766)
                        -- (9.99989, 0.872907);

                    \draw[
                        plotIII, join=round, thick, hatchborder
                    ]
                        (-5.95224, 5.99957)
                        -- (0.291645, 2.87854)
                        -- (0.275445, 2.88637)
                        -- (3.18997, 1.91242)
                        -- (3.19887, 1.91232)
                        -- (9.99989, 1.91201);
                              
                    \draw[
                        plotIII, join=round, thick, densely dashed
                    ]
                        (-8.58342, 5.9998)
                        -- (5.64901, 1.79017)
                        -- (5.6502, 1.78936)
                        -- (10.0005, 1.93105);
                    
                    \draw[
                        plotIV, join=round, thick, hatchborder
                    ]
                        (-4.975, 5.99993)
                        -- (0.668122, 3.17853)
                        -- (0.718448, 3.15709)
                        -- (3.53839, 2.21682)
                        -- (10, 2.21654);
                                  
                    \draw[
                        plotIV, join=round, thick, densely dashed
                    ]
                        (-9.25112, 6.00055)
                        -- (6.17655, 2.3673)
                        -- (6.1867, 2.36605)
                        -- (10, 3.08739);
                    
                    \draw[
                        plotV, join=round, thick, hatchborder
                    ]
                        (-4.23312, 5.99979)
                        -- (0.907885, 3.42967)
                        -- (0.97265, 3.40197)
                        -- (3.8153, 2.45432)
                        -- (3.83135, 2.45368)
                        -- (10.0001, 2.45369);
                             
                    \draw[
                        plotV, join=round, thick, densely dashed
                    ]  
                        (-10.0002, 5.65137)
                        -- (6.85547, 2.98035)
                        -- (6.86942, 2.9802)
                        -- (9.99945, 4.07537);

                    \filldraw [black] (-0.4, 4.3)circle[radius=1pt];
                    \draw[black, fill=black, fill opacity=0.3] (-1,4.2) rectangle (0.2,4.4);

                    \legend{$[12]$,$\lambda=4$,$\lambda=8$,$\lambda=16$,$\lambda=24$,$\lambda=32$,NLO,NNLO,,,,,,,};
                \end{axis}
            \end{tikzpicture}
        }
    \end{minipage}
    \begin{minipage}[t]{0.35\textwidth}
        \caption[Two-flavour two-derivative bounds on $\bar l_1,\bar l_2$.]{
            Two-flavour two-derivative NLO bounds on $\bar l_1,\bar l_2$ for various $\lambda$, as indicated in the legend, along with the Manohar--Mateu bound \cite{Manohar:2008tc}.
            Each set of constraints is represented as a line with one side hatched; the hatched side is excluded by the constraints, and the other side corresponds to $\sat(\Omega)$.
            The  experimentally measured reference point [\mbox{$\bar l_1=-0.4(6)$}, \mbox{$\bar l_2=4.3(1)$}] is drawn as a dot with an uncertainty region around it.
            For each bound, the direct NNLO counterpart (i.e.\ using the same $\lambda,s,t$ and $a_J$) is drawn as a dashed outline.}
        \label{fig:2flav-NLO}
    \end{minipage}
\end{figure}

In this section, we present a selection of our results; a larger selection can be found in \rcite{Alvarez:2021kpq}.
The simplest case is two-flavour \chpt\ at NLO, where there are only two LECs ($\bar l_1$ and $\bar l_2$) and constraints are available from Manohar \&\ Mateu \cite{Manohar:2008tc}.
We reproduce their bounds in \cref{fig:2flav-NLO}, along with our own.
The basic ($\lambda=4$) version of our constraints provides only marginal improvements on the Manohar--Mateu bounds, and in order to reach close to the experimental reference value \cite{Bijnens:2014lea}, very aggressive integration is needed, too close to the Chivukula--Dugan--Golden bound ($\lambda\sim35$) to be taken seriously.

\begin{figure}[hbtp]
    \centering
    \begin{tikzpicture}
        \begin{axis}[scale=.9,
            ylabel={$\bar l_2$}, xlabel={$\bar l_1$},
            xmin=-12.5,xmax=13.5, ymin=.5, ymax=9.5,
            legend pos=north east, legend cell align=left,
            legend image code/.code={
                \draw (0cm,0cm) -- (0.6cm,0cm);
            },
            legend style={font=\footnotesize}]
        
            \addlegendimage{plotI, thick, hatchborder};
            \addlegendimage{plotIII, thick, hatchborder};
            \addlegendimage{plotV, thick, hatchborder};
            \addlegendimage{plotgray, thick, hatchborder};
            \addlegendimage{plotgray, thick, densely dashed, hatchborder};
            \addlegendimage{plotgray, thick, densely dashed};
            
            \draw[
                plotI, join=round, thick, densely dashed
            ]
                (-12.0008,7.9626)
                -- (-0.324077,2.12455)
                -- (1.99952, 1.35006)
                -- (11.9996,1.35006);
            
            \draw[
                plotI, join=round, thick, 
                postaction={decorate,draw,thin}, invhatchborder
            ]
                (11.9995, 1.52292)
                -- (1.51172, 2.23894)
                -- (-1.31549, 2.69241)
                -- (-11.9993, 6.34788);

            \draw[
                plotI, join=round, thick, densely dashed, invhatchborder
            ]
                (-2.37543, 9.00002)
                -- (5.38005, 5.32254)
                -- (12.0004, 2.06887);

            
            \draw[
                plotIII, join=round, thick, densely dashed
            ]
                (-11.9996, 8.33135)
                -- (-0.202206, 2.43291)
                -- (-0.0944658, 2.38661)
                -- (2.52571, 1.51313)
                -- (2.54915, 1.51135)
                -- (12.0005, 1.51186);
            
            \draw[
                plotIII, join=round, thick, invhatchborder
            ]
                (11.9995, 1.77469)
                -- (2.55997, 2.51404)
                -- (1.18698, 2.66546)
                -- (0.238266, 2.79402)
                -- (-1.32693, 3.20141)
                -- (-10.8474, 6.39585)
                -- (-11.0738, 6.47263)
                -- (-11.2926, 6.54763)
                -- (-11.5254, 6.62962)
                -- (-11.8298, 6.73578)
                -- (-11.9995, 6.79618);

            \draw[
                plotIII, join=round, thick, densely dashed, invhatchborder
            ]
                (-11.3214, 9.00016)  
                -- (-10.9932, 8.95568)  
                -- (-9.91955, 8.77647)  
                -- (-8.71913, 8.54577)  
                -- (-7.37468, 8.25707)  
                -- (3.94601, 5.62959)  
                -- (12.0003, 1.4603)  
                (-6.04257, 1.00015)  
                -- (-7.76656, 2.83051)  
                -- (-8.07238, 3.18279)  
                --(-12.0005, 7.94612);  

            \draw[
                plotV, join=round, thick, densely dashed
            ]
                (-11.9999, 8.71883)
                -- (0.0317066, 2.70276)
                -- (0.0618442, 2.68807)
                -- (2.94007, 1.72857)
                -- (12.0002, 1.72886);
                
            \draw[
                plotV, join=round, thick, hatchborder
            ]
                (-12.0002, 7.41195)  
                -- (-6.3951, 5.328)  
                -- (-6.23606, 5.26938)  
                -- (-6.0821, 5.21449)  
                -- (-5.95212, 5.16781)  
                -- (-5.83237, 5.12652)  
                -- (-5.70553, 5.08279)  
                -- (-5.53063, 5.02404)  
                -- (-5.36457, 4.96969)  
                -- (-5.2976, 4.9482)  
                -- (-5.25458, 4.93383)  
                -- (-5.16758, 4.90607)  
                -- (-1.2058, 3.64465)  
                -- (0.524492, 3.24422)  
                -- (12.0002, 2.15303);  

            \draw[
                plotV, join=round, thick, densely dashed, hatchborder
            ]
                (11.6441, 0.999967)  
                -- (2.09598, 5.68563)  
                -- (-6.23222, 6.54122)  
                -- (-6.10807, 6.05474)  
                -- (-3.79306, 3.04983)  
                -- (-3.68259, 2.91218)  
                -- (-3.37137, 2.54966)  
                -- (-3.04907, 2.20042)  
                -- (-2.65723, 1.80026)  
                -- (-2.18061, 1.33768)  
                -- (-1.81728, 1.00001);  

            \filldraw [black] (-0.4, 4.3)circle[radius=1pt];
            \draw[black, fill=black, fill opacity=0.3] (-1,4.2) rectangle (0.2,4.4);
            
            \legend{$\lambda=4$,$\lambda=8$,$\lambda=12$,$k=2$,$k=4$,NLO,,,,,,,,,,};
                
        \end{axis}
    \end{tikzpicture}
    \begin{tikzpicture}
        \begin{axis}[scale=.9,
            xmin=-1100, xmax=800, ymin=-9, ymax=9,
            xlabel={$\bar l_3$},ylabel={$\bar l_4$},
            legend pos=south east, legend cell align=left,
            legend image code/.code={
                \draw (0cm,0cm) -- (0.6cm,0cm);
            },
            legend style={font=\footnotesize}]
        
            \addlegendimage{plotI, very thick, hatchborder};
            \addlegendimage{plotIII, very thick, hatchborder};
            \addlegendimage{plotV, very thick, hatchborder};
            \addlegendimage{plotgray, very thick, hatchborder};
            \addlegendimage{plotgray, very thick, densely dashed, hatchborder};
            
            \draw[
                plotI, join=round, thick, hatchborder
            ]  
                (-364.915, 8.00053)
                -- (-425.543, 6.36676)
                -- (-594.662, -6.54193)
                -- (-262.132, -3.51291)
                -- (-262.139, -3.51301)
                -- (523.714, 7.99791);

            \draw[
                plotI, join=round, thick, densely dashed, invhatchborder
            ]  
                (461.999, 7.998)
                -- (-593.287, -7.99831)
                (-710.014, -8.00266)
                -- (-492.186, -4.0817)
                -- (180.777, 8.00014);

            \draw[
                plotIII, join=round, thick, invhatchborder
            ]  
                (750.018, 4.86686)
                -- (-407.473, -4.07228)
                -- (-666.692, 8.0011);

            \draw[
                plotIII, join=round, thick, densely dashed, invhatchborder
            ]  
                (446.971, 7.9979)
                -- (-639.382, -7.99947)
                (-672.821, -8.00065)
                -- (-10.2216, 7.99947);

            \draw[
                plotV, join=round, thick, hatchborder
            ] 
                (-1000, 2.27101)
                -- (-349.385, -1.86502)
                -- (-205.514, -1.27891)
                -- (-205.515, -1.27892)
                -- (750.475, 5.23081);

            \draw[
                plotV, join=round, thick, densely dashed, invhatchborder
            ]  
                (337.902, 8.00054)
                -- (-502.897, -5.39663)
                -- (-502.896, -5.39661)
                -- (-272.846, 8.00031);
                
            \filldraw [black] (2.9, 4.4)circle[radius=1pt];
            \draw[black, fill=black, fill opacity=0.3] (0.5,4.2) rectangle (5.3,4.6);

            \legend{$\lambda=4$,$\lambda=8$,$\lambda=12$,$k=2$,$k=4$,,,,,,,,,,};
                
        \end{axis}
    \end{tikzpicture}
    \caption{NNLO bounds on $\bar l_1,\bar l_2$ (left) and $\bar l_3,\bar l_4$ (right), displayed similarly to \cref{fig:2flav-NLO}.
    In each plot, the LECs not shown have been fixed to their reference values.
    For comparison, the corresponding NLO bounds shown in \cref{fig:2flav-NLO} are drawn as a dashed outline.}
    \label{fig:2flav-NNLO}
\end{figure}
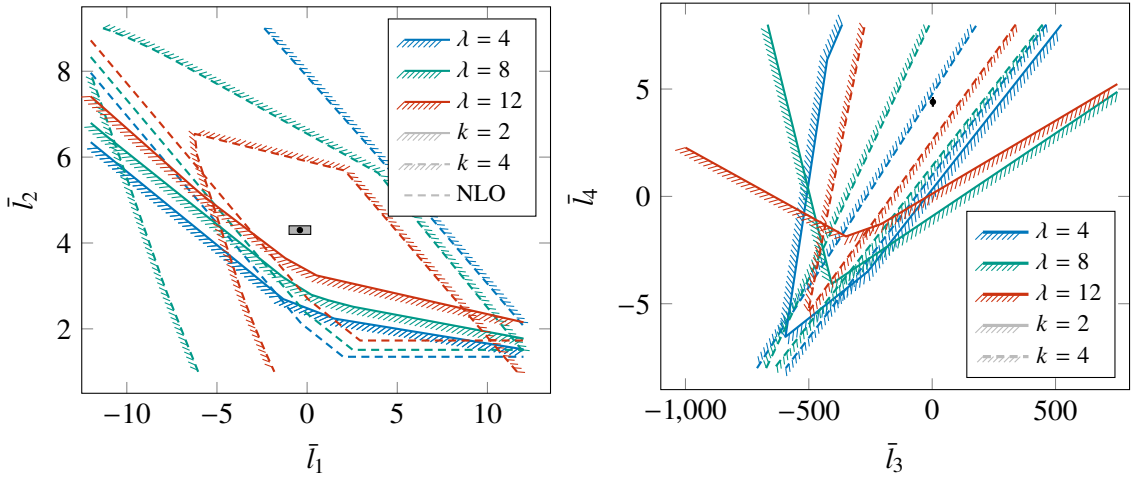

The bounds change significantly when the NNLO amplitude is used, even when just considering the LECs that also feature at NLO.
Two more NLO LECs ($\bar l_3$ and $\bar l_4$) enter the NNLO amplitude, along with four linear combinations of the NNLO LECs (not shown here).
\Cref{fig:2flav-NNLO} shows the bounds on the $\bar l_i$; there, also $k\geq 2$ can yield nontrivial bounds, although only $k=4$ yields useful ones.
With only modest integration, the allowed region in the $\bar l_1$--$\bar l_2$ plane becomes finite, albeit still not close to the experimental uncertainty.
Note that the bounds on $\bar l_3$ are extremely weak, since the coefficient of $\bar l_3$ in the amplitude is very small.

\begin{figure}[hbtp]
    \centering
    \begin{tikzpicture}
        \begin{axis}[
                scale=1.5,
                view={-250}{20}, zlabel={$10^{3} L_3^r$}, ylabel={$10^{3} L_2^r$}, xlabel={$10^{3} L_1^r$},
                colormap/temp, point meta min=0, point meta max=1000, colorbar,
                colorbar style={
                    ymin=0, ytick={200,400,600,800,1000}, yticklabels={0.2,0.4,0.6,0.8,1},ylabel={$10^3/\rho$}
                },
                zmin=-6.5,zmax=5,ymin=-1.32,ymax=3
            ]
            
            \addplot3 [
                patch, 
                patch table with point meta={\resultpath/NLO/SU3/M0.135/D2/lam4/visualisation/constr_0_table.dat}, 
                opacity=0] 
                    table {\resultpath/NLO/SU3/M0.135/D2/lam4/visualisation/constr_0_coords.dat};
        
            \coordinate (p0) at (-1.11889, -1.29675, 1.59467);
            \coordinate (p1) at (-1.54861, -1.24364, 2.35869);
            \coordinate (p2) at (-1.59259, -1.29691, 2.9321);
            \coordinate (p3) at (-1.76025, -1.28208, 3.28219);
            \coordinate (p4) at (-0.621654, -0.798673, -1.09356);
            \coordinate (p5) at (-0.636225, -0.799703, -1.06239);
            \coordinate (p6) at (-0.643296, -0.801226, -1.04428);
            \coordinate (p7) at (-0.568512, -0.834015, -1.09357);
            \coordinate (p8) at (-0.807803, -1.16889, 0.341358);
            \coordinate (p9) at (-1.3063, -1.08041, 1.04951);
            \coordinate (p10) at (-0.705188, -1.29519, 0.62573);
            \coordinate (p11) at (-0.684255, -1.27045, 0.515106);
            \coordinate (p12) at (4.99988, -1.29497, 0.625895);
            \coordinate (p13) at (4.99996, -1.27021, 0.515083);
            \coordinate (p14) at (-1.96597, -1.0934, 2.48368);
            \coordinate (p15) at (-1.98905, -1.08617, 2.50982);
            \coordinate (p16) at (-1.8916, -1.10663, 2.35879);
            \coordinate (p17) at (-1.86054, -1.10882, 2.29334);
            \coordinate (p18) at (-0.511848, -0.959599, -0.750789);
            \coordinate (p19) at (-0.494857, -0.843775, -1.15962);
            \coordinate (p20) at (5.00016, -0.959788, -0.750848);
            \coordinate (p21) at (4.99977, -0.843698, -1.15943);
            \coordinate (p22) at (-0.497222, -0.83443, -1.18744);
            \coordinate (p23) at (-0.530819, -0.82107, -1.18746);
            \coordinate (p24) at (-0.530684, -0.820945, -1.18766);
            \coordinate (p25) at (-0.497085, -0.834459, -1.18764);
            \coordinate (p26) at (-0.545701, -0.737866, -1.36722);
            \coordinate (p27) at (-4.67801, 0.210575, 4.99997);
            \coordinate (p28) at (0.30984, 0.973142, -6.50012);
            \coordinate (p29) at (-1.71681, 3.00001, -6.49989);
            \coordinate (p30) at (-4.99978, 3.00022, 0.065719);
            \coordinate (p31) at (-4.99995, 0.532897, 4.99995);
            \coordinate (p32) at (-0.529048, -0.797504, -1.24792);
            \coordinate (p33) at (0.3463, 0.953245, -6.4997);
            \coordinate (p34) at (0.388267, 0.936131, -6.50029);
            \coordinate (p35) at (-0.739558, -1.31531, 0.73855);
            \coordinate (p36) at (-0.735343, -1.31534, 0.732997);
            \coordinate (p37) at (-0.784907, -1.2971, 0.738527);
            \coordinate (p38) at (-0.521722, -0.996399, -0.614914);
            \coordinate (p39) at (-0.563218, -1.06996, -0.314218);
            \coordinate (p40) at (5.00021, 0.935992, -6.49994);
            \coordinate (p41) at (-0.864755, -1.31535, 1.06044);
            \coordinate (p42) at (-1.0363, -1.31537, 1.54455);
            \coordinate (p43) at (-2.18779, -1.31537, 4.99987);
            \coordinate (p44) at (4.9999, -1.31538, 0.733144);
            \coordinate (p45) at (5.00013, -1.31548, 5.00013);
            \coordinate (p46) at (4.99991, -1.07016, -0.314185);
            \coordinate (p47) at (-1.76702, -1.11962, 2.12156);
            \coordinate (p48) at (-4.6729, 0.206155, 5.00018);
            \coordinate (p49) at (-3.64928, -0.671204, 4.99979);
            \coordinate (p50) at (-2.8049, -1.09371, 4.99992);
            \coordinate (p51) at (-2.33264, -1.28244, 5.00004);
            \coordinate (p52) at (5.00021, -0.996242, -0.614692);
            \coordinate (p53) at (-1.56114, -1.30559, 2.93205);
            \coordinate (p54) at (-2.25066, -1.30598, 5.00014);
            
            \draw[color of colormap={418.55/1.04679}, facet]
                (p42) -- (p53) -- (p54) -- (p43) -- cycle;
            \draw[color of colormap={421.816/1.04679}, facet]
                (p2) -- (p3) -- (p51) -- (p54) -- (p53) -- cycle;
            \draw[color of colormap={422.773/1.04679}, facet]
                (p41) -- (p42) -- (p43) -- (p45) -- (p44) -- (p36) -- (p35) -- cycle;
            \draw[color of colormap={428.386/1.04679}, facet]
                (p1) -- (p3) -- (p51) -- (p50) -- (p14) -- (p16) -- cycle;
            \draw[color of colormap={429.006/1.04679}, facet]
                (p2) -- (p0) -- (p41) -- (p42) -- (p53) -- cycle;
            \draw[color of colormap={429.671/1.04679}, facet]
                (p15) -- (p14) -- (p50) -- (p49) -- cycle;
            \draw[color of colormap={432.842/1.04679}, facet]
                (p0) -- (p2) -- (p3) -- (p1) -- cycle;
            \draw[color of colormap={454.168/1.04679}, facet]
                (p0) -- (p1) -- (p16) -- (p17) -- (p47) -- (p37) -- (p35) -- (p41) -- cycle;
            \draw[color of colormap={492.472/1.04679}, facet]
                (p14) -- (p16) -- (p17) -- (p15) -- cycle;
            \draw[color of colormap={557.832/1.04679}, facet]
                (p8) -- (p9) -- (p47) -- (p37) -- cycle;
            \draw[color of colormap={605.724/1.04679}, facet]
                (p4) -- (p7) -- (p8) -- (p9) -- (p6) -- (p5) -- cycle;
            \draw[color of colormap={617.721/1.04679}, facet]
                (p6) -- (p9) -- (p47) -- (p17) -- (p15) -- (p49) -- (p48) -- cycle;
            \draw[color of colormap={638.545/1.04679}, facet]
                (p6) -- (p5) -- (p27) -- (p48) -- cycle;
            \draw[color of colormap={643.44/1.04679}, facet]
                (p35) -- (p37) -- (p8) -- (p7) -- (p23) -- (p22) -- (p19) -- (p18) -- (p38) -- (p39) -- (p11) -- (p10) -- (p36) -- cycle;
            \draw[color of colormap={670.601/1.04679}, facet]
                (p4) -- (p7) -- (p23) -- (p24) -- (p32) -- (p26) -- cycle;
            \draw[color of colormap={673.063/1.04679}, facet]
                (p10) -- (p36) -- (p44) -- (p12) -- cycle;
            \draw[color of colormap={676.548/1.04679}, facet]
                (p4) -- (p5) -- (p27) -- (p31) -- (p30) -- (p29) -- (p28) -- (p26) -- cycle;
            \draw[color of colormap={736.505/1.04679}, facet]
                (p32) -- (p26) -- (p28) -- (p33) -- cycle;
            \draw[color of colormap={740.479/1.04679}, facet]
                (p22) -- (p23) -- (p24) -- (p25) -- cycle;
            \draw[color of colormap={759.34/1.04679}, facet]
                (p10) -- (p11) -- (p13) -- (p12) -- cycle;
            \draw[color of colormap={779.273/1.04679}, facet]
                (p32) -- (p24) -- (p25) -- (p34) -- (p33) -- cycle;
            \draw[color of colormap={808.294/1.04679}, facet]
                (p11) -- (p39) -- (p46) -- (p13) -- cycle;
            \draw[color of colormap={815.898/1.04679}, facet]
                (p38) -- (p39) -- (p46) -- (p52) -- cycle;
            \draw[color of colormap={879.205/1.04679}, facet]
                (p18) -- (p38) -- (p52) -- (p20) -- cycle;
            \draw[color of colormap={911.721/1.04679}, facet]
                (p18) -- (p19) -- (p21) -- (p20) -- cycle;
            \draw[color of colormap={1046.79/1.04679}, facet]
                (p19) -- (p22) -- (p25) -- (p34) -- (p40) -- (p21) -- cycle;
            
            \def\xfid{1.11} \def\yfid{1.05}  \def\zfid{-3.82}
            \def\xmin{-5}   \def\ymin{-1.32} \def\zmin{-6.5}
            \def\xmax{5}   \def\ymax{3}   \def\zmax{5}
            
            \coordinate (fid) at (\xfid,\yfid,\zfid);
            
            \draw[gray, fill, opacity=.5]
                (p40) -- (p21) -- (p20) -- (p52) -- (p46) -- (p13) -- (p12) -- (p44) -- (p45) -- (\xmax,\ymin,\zmin) --  cycle;
            \draw[gray, fill, opacity=.5]
                (p29) -- (p30) -- (\xmin,\ymax,\zmin) -- cycle;
            \draw[gray, fill, opacity=.5]
                (p31) -- (p49) -- (p50) -- (p51) -- (p43) -- (\xmin,\ymin,\zmax) -- cycle;
            
            \draw[surface to fid]
                (1.11, 0.143718, -4.12209) -- (fid);
                
            \draw[axes to fid]
                (fid) -- (\xfid,\yfid,\zmax) -- (\xfid,\ymin,\zmax) (\xfid,\yfid,\zmax) -- (\xmin,\yfid,\zmax)
                (fid) -- (\xfid,\ymax,\zfid) -- (\xfid,\ymax,\zmin) (\xfid,\ymax,\zfid) -- (\xmin,\ymax,\zfid)
                (fid) -- (\xmax,\yfid,\zfid) -- (\xmax,\ymin,\zfid) (\xmax,\yfid,\zfid) -- (\xmax,\yfid,\zmin)
                (fid) -- (\xfid,\yfid,\zmin);
                
            \filldraw [black] (fid) circle[radius=1pt] {};

            \draw[black, fill=black, fill opacity=0.5, join=round] 
                (1.21,1.22,-3.52) -- (1.01,1.22,-3.52) -- (1.01,0.88,-3.52) -- (1.21,0.88,-3.52) -- cycle;
            \draw[black, fill=black, fill opacity=0.5, join=round] 
                (1.21,1.22,-3.52) -- (1.21,0.88,-3.52) -- (1.21,0.88,-4.12) -- (1.21,1.22,-4.12) -- cycle;
            \draw[black, fill=black, fill opacity=0.5, join=round] 
                (1.21,1.22,-3.52) -- (1.21,1.22,-4.12) -- (1.01,1.22,-4.12) -- (1.01,1.22,-3.52) -- cycle;
            
        \end{axis}
    \end{tikzpicture}
    \caption{Three-flavour two-derivative NLO bounds on $L_1^r$, $L_2^r$ and $L_3^r$.
        The space \emph{outside} $\sat(\Omega)$ is shown as a gray solid, with the empty space containing the reference point [\mbox{$L_1^r = 0.00111(10)$}, \mbox{$L_2^r = 0.00105(17)$}, \mbox{$L_3^r = -0.00382(30)$}] being a part of $\sat(\Omega)$.
        The constraint surfaces are coloured according to the orthogonal distance to the reference point, denoted $\rho$, and the orthogonal line (which does not appear as such due to different axis scales) from the point to the surface is drawn whenever possible.
        Dotted lines are drawn parallel to the axes to clarify the reference point's position in space.}
    \label{fig:3flav-NLO}
\end{figure}
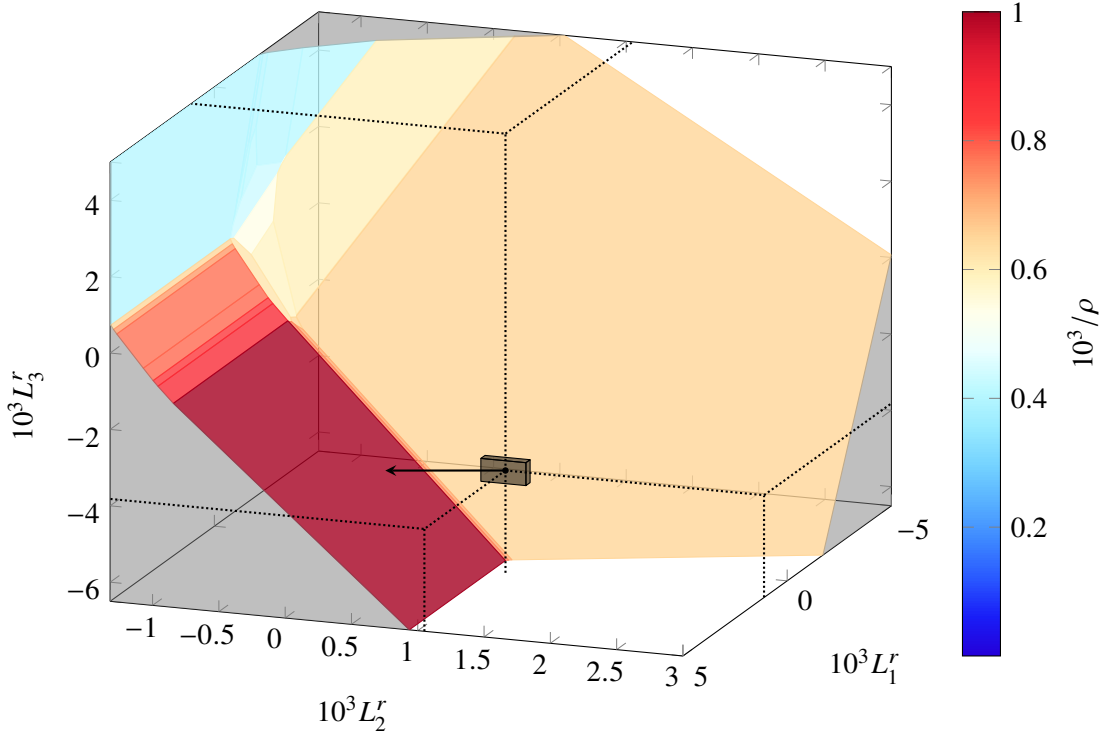

With three flavours, the LECs entering at NLO are $L_1^r$, $L_2^r$ and $L_3^r$ (the `$r$' indicating a different renormalization convention than the bar on $\bar l_i$).
Thus, we use a three-dimensional visualization of the bounds, shown in \cref{fig:3flav-NLO}, which is generated using \cref{prop:construct}.

Qualitatively, the bounds in \cref{fig:3flav-NLO} are similar to those in \cref{fig:2flav-NLO}, although here there is no earlier result to compare to; Mateu \cite{Mateu:2008gv} uses realistic meson masses.
The bounds depend strongly on the choice of $M$ (only $M=M_\pi$ is shown here) and are significantly more sensitive to integration (not shown) than their two-flavour counterparts. 
This sensitivity to the details of the isospin limit reduces the applicability of our $(n>2)$-flavour bounds; unfortunately, the $2\to2$ scattering amplitude with realistic meson masses is not known at NNLO, just NLO \cite{GomezNicola:2001as}.

At NNLO, in addition to $L_{1,2,3}^r$ whose NNLO bounds are shown in \cref{fig:3flav-NNLO}, four more NLO LECs ($L_{4,5,6,8}^r$) and five linear combinations of NNLO LECs ($\Xi_{1,2,3},\Gamma_3,\Delta_3$; these are defined in \rcite{Alvarez:2021kpq}) appear in the amplitude.
The former (not shown) are bounded similarly to $\bar l_{3,4}$, and the reference point is excluded already at $\lambda=4.5$.
Several of the latter are constrained to finite ranges which exclude the reference point even without integration, as shown in \cref{fig:3flav-GDX}, although it must be kept in mind that the NNLO LECs are only roughly estimated in \rcite{Bijnens:2014lea}.
The bounds are quite sensitive to the values of the other LECs, which are fixed to their reference values to produce these figures, but no value of the NLO LECs within their experimental uncertainties allow the reference point to satisfy the bounds in \cref{fig:3flav-GDX}.

\begin{figure}[hbtp]
    \hspace{-2cm}
    \begin{tikzpicture}
        \begin{axis}[
                scale=.9,
                view={-250}{20}, zlabel={$10^{3} L_3^r$}, ylabel={$10^{3} L_2^r$}, xlabel={$10^{3} L_1^r$},
                colormap/temp, point meta min=0, point meta max=1600,
                zmin=-8,zmax=6, ymin=-1.9,ymax=3
            ]

            \def\xfid{1.11} \def\yfid{1.05}  \def\zfid{-3.82}
            \def\xmin{-5}   \def\ymin{-1.9} \def\zmin{-8}
            \def\xmax{5}   \def\ymax{3}   \def\zmax{6}
            
            \addplot3 [
                patch, 
                patch table with point meta={\resultpath/NNLO/SU3/M0.135/D2/lam4/visualisation/L123/constr_0_table.dat}, 
                opacity=0] 
                    table {\resultpath/NNLO/SU3/M0.135/D2/lam4/visualisation/L123/constr_0_coords.dat};
                    
            \coordinate (p0) at (1.57844, -1.7655, 6.00026);
            \coordinate (p1) at (1.2159, -1.72147, 6.00002);
            \coordinate (p2) at (2.25844, -1.72882, 4.24211);
            \coordinate (p3) at (2.36965, -1.75887, 4.47762);
            \coordinate (p4) at (2.3303, -1.77081, 4.72901);
            \coordinate (p5) at (-4.76306, -0.952611, 5.99975);
            \coordinate (p6) at (-4.76307, -0.952615, 5.99977);
            \coordinate (p7) at (-1.18195, -0.953093, -0.39388);
            \coordinate (p8) at (-1.18196, -0.953094, -0.394037);
            \coordinate (p9) at (2.19138, -1.82925, 6.00006);
            \coordinate (p10) at (4.99966, -1.59952, 4.93983);
            \coordinate (p11) at (4.9999, -1.64083, 5.99988);
            \coordinate (p12) at (-5.00006, 1.72385, -2.11692);
            \coordinate (p13) at (-2.56274, 2.20234, -7.99999);
            \coordinate (p14) at (-0.999996, -0.782769, -1.26271);
            \coordinate (p15) at (-0.429337, 1.00935, -7.9997);
            \coordinate (p16) at (-5.00009, -0.819868, 5.9998);
            \coordinate (p17) at (-0.00489809, -0.827492, -1.13497);
            \coordinate (p18) at (2.94048, -0.57118, -2.21579);
            \coordinate (p19) at (4.99984, -0.0208327, -4.36093);
            \coordinate (p20) at (4.99971, 0.950068, -7.99973);
            \coordinate (p21) at (1.62316, -1.54497, 2.82116);
            \coordinate (p22) at (-3.65563, 2.83321, -7.9997);
            \coordinate (p23) at (-5.00008, 2.99988, -6.00814);
            \coordinate (p24) at (-5.00003, 2.56848, -4.75341);
            \coordinate (p25) at (-3.91873, 3.00038, -8.00007);
            \coordinate (p26) at (-0.431624, -1.09548, 0.244735);
            
            \draw[color of colormap={522.938/1.62448}, facet]
                (p0) -- (p4) -- (p3) -- (p2) -- (p1) -- cycle;
            \draw[color of colormap={530.667/1.62448}, facet]
                (p5) -- (p7) -- (p8) -- (p6) -- cycle;
            \draw[color of colormap={494.84/1.62448}, facet]
                (p0) -- (p4) -- (p9) -- cycle;
            \draw[color of colormap={1904.54/1.62448}, facet]
                (p3) -- (p4) -- (p9) -- (p11) -- (p10) -- cycle;
            \draw[color of colormap={556.599/1.62448}, facet]
                (p12) -- (p13) -- (p15) -- (p14) -- (p8) -- (p6) -- (p16) -- cycle;
            \draw[color of colormap={908.402/1.62448}, facet]
                (p17) -- (p14) -- (p15) -- (p20) -- (p19) -- (p18) -- cycle;
            \draw[color of colormap={1624.48/1.62448}, facet]
                (p2) -- (p3) -- (p10) -- (p19) -- (p18) -- (p21) -- cycle;
            \draw[color of colormap={477.304/1.62448}, facet]
                (p22) -- (p24) -- (p23) -- (p25) -- cycle;
            \draw[color of colormap={1250.87/1.62448}, facet]
                (p26) -- (p17) -- (p18) -- (p21) -- cycle;
            \draw[color of colormap={531.15/1.62448}, facet]
                (p1) -- (p2) -- (p21) -- (p26) -- (p7) -- (p5) -- cycle;
            \draw[color of colormap={537.953/1.62448}, facet]
                (p24) -- (p22) -- (p13) -- (p12) -- cycle;
            \draw[color of colormap={897.881/1.62448}, facet]
                (p7) -- (p8) -- (p14) -- (p17) -- (p26) -- cycle;
                
            \coordinate (fid) at (\xfid,\yfid,\zfid);
            
            \draw[gray, fill, opacity=.5]
                (p20) -- (p19) -- (p10) -- (p11) -- (\xmax,\ymin,\zmax) -- (\xmax,\ymin,\zmin) -- cycle;
            \draw[gray, fill, opacity=.5]
                (p25) -- (p23) -- (\xmin,\ymax,\zmin) -- cycle;
            \draw[gray, fill, opacity=.5]
                (p16) -- (p6) -- (p1) -- (p0) -- (p9) -- (p11) -- (\xmax,\ymin,\zmax) -- (\xmin,\ymin,\zmax) -- cycle;
            
            \draw[surface to fid]
                (1.09815, -0.0449969, -4.11219) -- (fid);
                
            \draw[axes to fid]
                    (fid) -- (\xfid,\yfid,\zmax) -- (\xfid,\ymin,\zmax) (\xfid,\yfid,\zmax) -- (\xmin,\yfid,\zmax)
                    (fid) -- (\xfid,\ymax,\zfid) -- (\xfid,\ymax,\zmin) (\xfid,\ymax,\zfid) -- (\xmin,\ymax,\zfid)
                    (fid) -- (\xmax,\yfid,\zfid) -- (\xmax,\ymin,\zfid) (\xmax,\yfid,\zfid) -- (\xmax,\yfid,\zmin)
                    (fid) -- (\xfid,\yfid,\zmin);
                
            \filldraw [black] (\xfid,\yfid,\zfid) circle[radius=1pt] {};

            \draw[black, fill=black, fill opacity=0.5, join=round] 
                (1.21,1.22,-3.52) -- (1.01,1.22,-3.52) -- (1.01,0.88,-3.52) -- (1.21,0.88,-3.52) -- cycle;
            \draw[black, fill=black, fill opacity=0.5, join=round] 
                (1.21,1.22,-3.52) -- (1.21,0.88,-3.52) -- (1.21,0.88,-4.12) -- (1.21,1.22,-4.12) -- cycle;
            \draw[black, fill=black, fill opacity=0.5, join=round] 
                (1.21,1.22,-3.52) -- (1.21,1.22,-4.12) -- (1.01,1.22,-4.12) -- (1.01,1.22,-3.52) -- cycle;
            
        \end{axis}
    \end{tikzpicture}
    \hspace{-.5cm}
    \begin{tikzpicture}
        \begin{axis}[
                scale=.9,
                view={-345}{20}, ylabel={$10^{3} L_2^r$}, xlabel={$10^{3} L_1^r$},
                colormap/temp, colorbar, point meta min=0, point meta max = 1600,
                colorbar style={
                    ymin=0, ytick={500,1000,1500}, yticklabels={0.5,1.0,1.5},ylabel={$10^3/\rho$}
                },
                zmin=-16,zmax=12
            ]
        
            \def\xfid{1.05} \def\yfid{1.11}  \def\zfid{-3.82}
            \def\xmin{-10}   \def\ymin{-6} \def\zmin{-16}
            \def\xmax{10}   \def\ymax{6}   \def\zmax{12}
        
            \addplot3 [
                patch, 
                patch table with point meta={\resultpath/NNLO/SU3/M0.135/D4/lam4/visualisation/L123/constr_0_table.dat}, 
                opacity=.1] 
                    table {\resultpath/NNLO/SU3/M0.135/D4/lam4/visualisation/L123/constr_0_coords.dat};
                    
            \coordinate (p0) at (4.75244, 5.52605, -15.9997);
            \coordinate (p1) at (10.0002, 1.42633, -16.0003);
            \coordinate (p2) at (9.85703, 1.24745, -14.4722);
            \coordinate (p3) at (9.99946, 1.15674, -14.5793);
            \coordinate (p4) at (2.01901, -6.00014, -9.73281);
            \coordinate (p5) at (2.12323, -5.02797, -11.2501);
            \coordinate (p6) at (6.54345, -5.99999, -16);
            \coordinate (p7) at (3.15302, -2.67213, -15.9998);
            \coordinate (p8) at (-2.0227, 6.00009, -1.37287);
            \coordinate (p9) at (9.99965, -6.00053, -4.12405);
            \coordinate (p10) at (-9.95016, 5.99983, 11.9997);
            \coordinate (p11) at (0.441773, -6.00038, 11.9994);
            \coordinate (p12) at (4.11651, 6.00076, -16.0003);
            \coordinate (p13) at (-7.5395, -5.99962, 11.6026);
            \coordinate (p14) at (-7.73757, -4.46538, 10.0166);
            \coordinate (p15) at (-10.0006, -1.57309, 12.0007);
            \coordinate (p16) at (-9.99944, 5.99991, 4.0588);
            \coordinate (p17) at (-8.81202, -4.0849, 12);
            \coordinate (p18) at (-0.950363, 6.00031, -16.0003);
            \coordinate (p19) at (-7.71091, -5.99944, 12.0004);

            \draw[color of colormap={272.137/1.62448}, facet]
                (p0) -- (p2) -- (p3) -- (p1) -- cycle;
            \draw[color of colormap={155.555/1.62448}, facet]
                (p4) -- (p5) -- (p7) -- (p6) -- cycle;
            \draw[color of colormap={562.134/1.62448}, facet]
                (p2) -- (p3) -- (p9) -- (p11) -- (p10) -- (p8) -- cycle;
            \draw[color of colormap={417.512/1.62448}, facet]
                (p2) -- (p0) -- (p12) -- (p8) -- cycle;
            \draw[color of colormap={216.021/1.62448}, facet]
                (p4) -- (p5) -- (p14) -- (p13) -- cycle;
            \draw[color of colormap={233.89/1.62448}, facet]
                (p5) -- (p14) -- (p17) -- (p15) -- (p16) -- (p18) -- (p7) -- cycle;
            \draw[color of colormap={210.352/1.62448}, facet]
                (p13) -- (p14) -- (p17) -- (p19) -- cycle;
                
            \coordinate (fid) at (\xfid,\yfid,\zfid);
            
            \draw[gray, fill, opacity=.5]
                (p6) -- (p4) -- (p13) -- (p19) -- (\xmin,\ymin,\zmax) -- (\xmin,\ymin,\zmin) -- cycle;
            \draw[gray, fill, opacity=.5]
                (p19) -- (p17) -- (p15) -- (\xmin,\ymin,\zmax) -- cycle;
            \draw[gray, fill, opacity=.5]
                (p10) -- (p11) -- (\xmax,\ymin,\zmax) -- (\xmax,\ymax,\zmax) -- cycle;
            \draw[gray, fill, opacity=.5]
                (p11) -- (p9) -- (\xmax,\ymin,\zmax) -- cycle;
            \draw[gray, fill, opacity=.5]
                (p9) -- (p3) -- (p1) -- (\xmax,\ymax,\zmin) -- (\xmax,\ymax,\zmax) -- (\xmax,\ymin,\zmax) -- cycle;
            
            \draw[surface to fid]
                (2.34947, 2.12336, -3.08526) -- (fid);
            \draw[surface to fid]
                (-2.54267, -0.678028, -5.46793) -- (fid);
                
            \draw[axes to fid]
                    (fid) -- (\xfid,\ymin,\zfid) -- (\xmin,\ymin,\zfid) (\xfid,\ymin,\zfid) -- (\xfid,\ymin,\zmin);
                
            \filldraw [black] (\xfid,\yfid,\zfid) circle[radius=1pt] {};

            \draw[black, fill=black, fill opacity=0.5, join=round] 
                (1.21, 1.22, -3.52) -- (1.01, 1.22, -3.52) -- (1.01, 0.88, -3.52) -- (1.21, 0.88, -3.52) -- cycle;
            \draw[black, fill=black, fill opacity=0.5, join=round] 
                (1.21, 1.22, -3.52) -- (1.21, 0.88, -3.52) -- (1.21, 0.88, -4.12) -- (1.21, 1.22, -4.12) -- cycle;
            \draw[black, fill=black, fill opacity=0.5, join=round] 
                (1.21, 0.88, -3.52) -- (1.21, 0.88, -4.12) -- (1.01, 0.88, -4.12) -- (1.01, 0.88, -3.52) -- cycle;
            
        \end{axis}
    \end{tikzpicture}
    \caption{
        NNLO bounds on $L_1^r,L_2^r$ and $L_3^r$, using two (left) and four (right) derivatives.
        Thus, the left figure is essentially the NNLO version of \cref{fig:3flav-NLO}. 
        In the four-derivative case, $\sat(\Omega)$ is actually finite: 
        It is a lentil-shaped body whose largest dimension is about two orders of magnitude larger than the region shown in the figure. Note that the axes have been rotated relative to the left figure in order to make the inside of $\sat(\Omega)$ reasonably visible.}
    \label{fig:3flav-NNLO}
\end{figure}
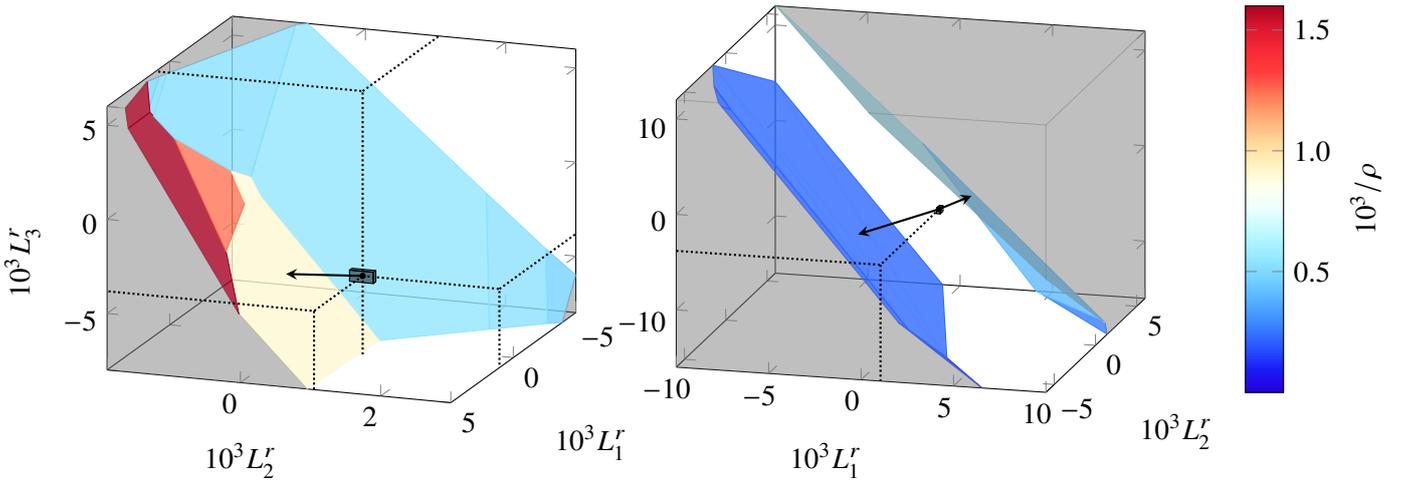

\begin{figure}[hbtp]
    \centering
    \begin{tikzpicture}
        \begin{axis}[
                view={170}{50}, xlabel={$10^{3} \Xi_1$}, ylabel={$10^{3} \Xi_2$}, zlabel={$10^{3} \Xi_3$},
                colormap/temp, colorbar, point meta min=0, point meta max = 720.041,
                colorbar style={
                    ymin=0, ytick={200,400,600}, yticklabels={0.2,0.4,0.6},ylabel={$10^3/\rho\left(\constr{\v\alpha_i}{c_i}\right)$}
                },
                xmin=-10,xmax=10,
                ymin=-10,ymax=10,
                zmin=-3, zmax=10,
            ]

            \addplot3 [patch, patch table with point meta={\tresultpath/onlyNNLO/SU3/M0.135/D2/lam4/visualisation/X123/constr_0_table.dat}, opacity=0] 
                    table {\tresultpath/onlyNNLO/SU3/M0.135/D2/lam4/visualisation/X123/constr_0_coords.dat};

            \coordinate (p0) at (-10.0007, -2.02727, 10.0007);
            \coordinate (p1) at (0.20276, 2.41665, -1.24568);
            \coordinate (p2) at (0.948376, 1.67097, -0.997299);
            \coordinate (p3) at (-9.99977, 2.90787, 5.06531);
            \coordinate (p4) at (-1.27077, 2.90784, -0.754765);
            \coordinate (p5) at (-2.29861, -7.16112, 10.0003);
            \coordinate (p6) at (7.5731, -7.16176, 3.41908);
            \coordinate (p7) at (0.202796, 9.99971, -1.24581);
            \coordinate (p8) at (0.948354, 10.0004, -0.99721);
            \coordinate (p9) at (0.948375, 1.67097, -0.997298);
            \coordinate (p10) at (0.948374, 1.67097, -0.997297);
            \coordinate (p11) at (0.202759, 2.41664, -1.24567);
            \coordinate (p12) at (10, -10, 7.47722);
            \coordinate (p13) at (6.21639, -9.99967, 9.99967);
            \coordinate (p14) at (9.99994, -8.77911, 5.03696);
            \coordinate (p15) at (-2.29861, -7.16109, 10.0002);
            \coordinate (p16) at (-9.9998, 2.90788, 5.06533);
            \coordinate (p17) at (-10, 10, 5.06535);
            \coordinate (p18) at (-1.27044, 2.90781, -0.75476);
            \coordinate (p19) at (-1.27065, 10.0004, -0.754607);
            \coordinate (p20) at (9.99979, 9.99979, 5.03668);
            \coordinate (p21) at (0.202759, 2.41664, -1.24567);

            \draw[plotIII, facet]
                (p0)-- (p5)-- (p6)-- (p2)-- (p1)-- (p4)-- (p3)
                -- cycle;
            \draw[color of colormap={720.41/0.72041}, facet]
                (p7)-- (p8)-- (p10)-- (p9)-- (p11)
                -- cycle;
            \draw[color of colormap={334.86/0.72041}, facet]
                (p12)-- (p13)-- (p15)-- (p5)-- (p14)
                -- cycle;
            \draw[color of colormap={588.467/0.72041}, facet]
                (p16)-- (p17)-- (p19)-- (p18)
                -- cycle;
            \draw[color of colormap={714.379/0.72041}, facet]
                (p6)-- (p5)-- (p14)
                -- cycle;
            \draw[color of colormap={714.379/0.72041}, facet]
                (p8)-- (p10)-- (p6)-- (p14)-- (p20)
                -- cycle;
            \draw[color of colormap={692.626/0.72041}, facet]
                (p7)-- (p11)-- (p21)-- (p18)-- (p19)
                -- cycle;
            
            \draw[gray, fill, opacity=.5]
                (p20) -- (p8) -- (p7) -- (p19) -- (p17) -- (\xmin,\ymax,\zmin) -- (\xmax,\ymax,\zmin) -- cycle;
            \draw[gray, fill, opacity=.5]
                (p20) -- (p14) -- (p12) -- (\xmax,\ymin,\zmin) -- (\xmax,\ymax,\zmin) -- cycle;
            \draw[gray, fill, opacity=.5]
                (p13) -- (p5) -- (p0) -- (\xmin,\ymin,\zmax) -- cycle;    
                
            \def\xfid{0.2908}   \def\yfid{0.336} \def\zfid{0.2468}
            \def\xmin{-10},\def\xmax{10},
            \def\ymin{-10},\def\ymax{10},
            \def\zmin{-3}, \def\zmax{10},
                        
            \draw[axes to fid] (\xfid,\ymax,\zfid) -- (\xfid,\yfid,\zfid);
            \draw[axes to fid] (\xfid,\ymax,\zmin) -- (\xfid,\ymax,\zfid) -- (\xmin,\ymax,\zfid);
            \draw[axes to fid] (\xfid,\yfid,\zmax) -- (\xfid,\yfid,\zfid);
            \draw[axes to fid] (\xfid,\ymin,\zmax) -- (\xfid,\yfid,\zmax) -- (\xmin,\yfid,\zmax);
            
            \filldraw [black] (\xfid,\yfid,\zfid) circle[radius=1pt];
                    
        \end{axis}
    \end{tikzpicture}
    
    \hspace{-1cm}
    \begin{tikzpicture}
        \begin{axis}[width=0.35\textwidth,
            xlabel={$10^{3}\Gamma_3$}, ylabel={$10^{3}\Delta_3$},
            xmin=-0.15,]

            \addplot [colormap={cc}{color=(red) color=(red)},
                    name path=lam4onlyNNLO, join=round,
                    patch, patch type=line, very thick, 
                    patch table={\tresultpath/onlyNNLO/SU3/M0.135/D2/lam4/visualisation/GD/constr_0_table.dat},
                    opacity=0
                ] 
                table {\tresultpath/onlyNNLO/SU3/M0.135/D2/lam4/visualisation/GD/constr_0_coords.dat};
                        
            \draw[plotI, thick, join=round, invhatchborder]
                (0.0804473, -0.856614) 
                -- (0.0655725, -0.142755) 
                -- (0.068524, -0.10257) 
                -- (0.0741833, -0.0639481) 
                -- (0.0818041, -0.0269408) 
                -- (0.090816, 0.00846321) 
                -- (0.100794, 0.0422917) 
                -- (0.111461, 0.074652) 
                -- (0.122551, 0.105522) 
                -- (0.133903, 0.135044) 
                -- (0.145355, 0.163142) 
                -- (0.156874, 0.190097) 
                -- (0.168363, 0.215888) 
                -- (0.179637, 0.240376) 
                -- (0.190810, 0.263845) 
                -- (0.201832, 0.286455) 
                -- (0.337266, 0.557333) 
                -- (0.337274, 0.557345) 
                -- (0.337289, -0.819992) 
                -- (0.190725, -0.846547) 
                -- cycle;

            \filldraw [black] (-0.10, -0.048) circle[radius=1pt];
                
        \end{axis}
    \end{tikzpicture}
    \begin{tikzpicture}
        \begin{axis}[width=0.35\textwidth,
            xlabel={$10^{3}\Gamma_3$}, ylabel={$10^{3}\Xi_4$},
            xmin=-0.15,]

            \addplot [colormap={cc}{color=(red) color=(red)},
                    name path=lam4onlyNNLO, join=round,
                    patch, patch type=line, very thick, 
                    patch table={\tresultpath/onlyNNLO/SU3/M0.135/D2/lam4/visualisation/GX/constr_0_table.dat},
                    opacity=0
                ] 
                table {\tresultpath/onlyNNLO/SU3/M0.135/D2/lam4/visualisation/GX/constr_0_coords.dat};
                
            \draw[plotI, thick, join=round, hatchborder]
                (0.0665345, 0.00129436) 
                -- (0.0884575, -0.0173436) 
                -- (0.341953, -0.165808) 
                -- (0.337595, 1.81261) 
                -- (0.109256, 1.2047) 
                -- (0.0713753, 1.04815) 
                -- (0.0346037, 0.875122) 
                -- (0.0346038, 0.0726519) 
                -- (0.0357747, 0.060867) 
                -- (0.0438476, 0.0338243) 
                -- (0.0525874, 0.0183143) 
                -- cycle;

            \filldraw [black] (-0.10, -0.008) circle[radius=1pt];
                
        \end{axis}
    \end{tikzpicture}
    \begin{tikzpicture}
        \begin{axis}[width=0.35\textwidth,
            xlabel={$10^{3}\Delta_3$}, ylabel={$10^{3}\Xi_4$},
            ymin=-0.05,]

            \addplot [colormap={cc}{color=(red) color=(red)},
                    name path=lam4onlyNNLO, join=round,
                    patch, patch type=line, very thick, 
                    patch table={\tresultpath/onlyNNLO/SU3/M0.135/D2/lam4/visualisation/DX/constr_0_table.dat},
                    opacity=0
                ] 
                table {\tresultpath/onlyNNLO/SU3/M0.135/D2/lam4/visualisation/DX/constr_0_coords.dat};
                
            \draw[plotI, thick, join=round, invhatchborder]
                (-0.325217, 0.141974) 
                -- (-0.327536, 0.130739) 
                -- (-0.333217, 0.119272) 
                -- (-0.343692, 0.107733) 
                -- (-0.361158, 0.0966078) 
                -- (-0.389057, 0.0865722) 
                -- (-0.43292, 0.0789008) 
                -- (-0.996461, 0.054738) 
                -- (-1.21992, 0.188876) 
                -- (-0.325218, 0.18882) 
                -- (-0.325219, 0.188821) 
                -- cycle;

            \filldraw [black] (-0.48, -0.008) circle[radius=1pt];
                
        \end{axis}
    \end{tikzpicture}
    \caption{NNLO two-derivative bounds on $\Xi_{1,2,3}$ (top) and $\Gamma_3,\Delta_3,\Xi_3$ (bottom).
        Two-dimensional slices are used since the finite $\sat(\Omega)$ makes three-dimensional representation difficult.
        The green constraint surface excludes the reference point [\mbox{$\Xi_1=2.9\cdot 10^{-4}$}, \mbox{$\Xi_2=3.4\cdot 10^{-4}$}, \mbox{$\Xi_3=2.5\cdot 10^{-4}$}, \mbox{$\Xi_4=-8\cdot 10^{-6}$}, \mbox{$\Gamma_3=-1.0\cdot 10^{-4}$}, \mbox{$\Delta_3=-4.8\cdot 10^{-5}$}], albeit by a very small amount not visible at the scale of the figure.
        No uncertainties are available on the estimates of the NNLO LECs \cite{Bijnens:2014lea}.}
    \label{fig:3flav-GDX}
\end{figure}

With four or more flavours, a few more LECs enter the amplitude, but their bounds (not shown) are qualitatively similar to those at three flavours.
The bounds gradually grow weaker as the number of flavours increases, and asymptotically approach triviality (i.e.\ being satisfied by all points) as $n\to\infty$, as can be deduced from the amplitude.
Many-flavour bounds are not readily interpreted due to the unphysicality of many-flavour \chpt, and care must be taken about perturbativity, since the Chivukula--Dugan--Golden bound scales as $1/n$.

\section{Summary and Outlook}
We present the first general-flavour NNLO bounds, albeit in the isospin limit, and present some generalizations of the Manohar--Mateu method, in particular in the treatment of the isospin decomposition coefficient $a_J$.
We also describe a new mathematical framework for managing large numbers of constraints in high-dimensional parameter spaces.
In the cases where previously derived bounds exist, our results provide some improvement, but do not come close to the experimental uncertainty without the use of hard-to-motivate amounts of integration.

Possible further development, besides refinement of our methods, would mostly require hitherto unknown amplitudes: NNLO beyond the isospin limit, or N$^3$LO, where an additional complication is that terms non-linear in the LECs appear, requiring generalization of the linear constraint framework.
Alternatively, bounds on the recently calculated NLO $2\to4$ amplitudes \cite{Bijnens:2021hpq,Bijnens:2022zsq} could be explored, although this would require generalization of the derivation of bounds.
Lastly, one can go beyond \chpt; to a large extent, these methods could be applied as-is to other EFTs, such as those used in beyond-the-Standard-Model research.

\acknowledgments
This work is supported in part by the Swedish Research Council grants contract numbers 2016-05996 and 2019-03779.


\providecommand{\href}[2]{#2}\begingroup\raggedright\endgroup


\end{document}